# UBIQUITOUS SCAFFOLD LEARNING ENVIRONMENT USING PROBLEM-BASED LEARNING TO ENHANCE PROBLEM-SOLVING SKILLS AND CONTEXT AWARENESS

Noppadon Phumeechanya[1] and Panita Wannapiroon, Ph.D.[2]

[1]Ph.D. Candidate, Department of Information and Communication Technology for Education, Faculty of Technical Education, King Mongkut's University of Technology North Bangkok, Thailand.
[2]Assistant Professor, Information and Communication Technology for Education Division, Faculty of Technical Education, King Mongkutt's University of Technology North Bangkok, Thailand.

## ABSTRACT

*The purpose of this research is to 1) design of an Ubiquitous Scaffold Learning Environment Using Problem-based Learning model to enhance problem-solving skills and context awareness, and 2) evaluate the developed model. The research procedures divide into two phases. The first phase is to design of Ubiquitous Scaffold Learning Environment Using Problem-based Learning model, and the second phase is to evaluate the developed model. The sample group in this study consists of five experts selected by purposive sampling method. Data were analyzed by arithmetic mean and standard deviation. The research findings are as follows: 1. The developed model consist of three components is 1) principles of ubiquitous learning environment (ULE), problem-based learning with scaffolding in ULE, problem solving skill and context awareness 2) objectives of the model are to enhance problem solving skill and context awareness and 3) Process of the instructional model 2. The experts agree Ubiquitous Scaffold Learning Environment Using Problem-based Learning model model is high level of appropriateness.*

## KEYWORDS

*Problem-based Learning, Ubiquitous Learning, Scaffolding, Problem-solving Skills, Context Awareness*

## 1. INTRODUCTION

The development of problem-solving skill is indispensable and it is also an important objective of the present education management. This is because it is necessary for the life-long learning and it corresponds to Partnership for 21st Century Skills [1], which requires that the learners must have development of problem-solving skill, one of the learning skills and innovations. The problem-solving skill can be developed by means of Problem-based learning (PBL) [2], in which the learners are encouraged to create new knowledge from the real-life problems. It is also a learner-centered instructional model. Referring to the learning process, any educational institute must encourage the students to learn about thinking process, and hold any activity to let the students learn from real experiences, think critically, take action and possess active learning. Thereby, the learning process that enables the students to have problem-solving skills is Problem-based Learning or PBL, a learning model based on the theories of Constructivism. In this model, the students generate new knowledge from the problems of the real world, which is considered a learning context to help the students receive skills of analytical thinking and problem-solving thinking[2]. According to National ICT Policy Framework 2011-2020, the education in Thailand is projected to step into "Smart Learning", which continually supports the application and development of ICT learning media for educational personnel. This will lead to a life long learning society[3].

During the Problem-based learning, the instructors must have enough skills to stimulate and advise the learners, and support them with relevant information sources. This kind of assistance is called Scaffolding [4]. It refers to the assistance in the form of support for learners to complete the tasks that they cannot do by themselves. When the learners begin to do the said tasks on their own, the said support will be gradually reduced until the learners can fully undertake the tasks by themselves.

Ubiquitous Learning (u-Learning) refers to a learning model in which the learners can learn anytime anywhere with the aid of portable computer technology and wireless communication; and thereby, the said learning must emphasize on context-awareness of the learners. The term "u-learning" stands for ubiquitous learning. The word 'ubiquitous' means everywhere. The combination of that with the word 'learning' denotes a learning model that allows learners to gain knowledge anywhere by using mobile computer technology and wireless communication as tools. The learning recognizes learner's context. The so-called ubiquitous learning environment (ULE) [5] is a setting that encourages pervasive study. Learning can happen anytime with a mobile computer mediating an access to knowledge sources. The right learning theory for ubiquitous learning environment is constructivism.

Ubiquitous Learning Environment (ULE) is a management of learning environment in which everything can be learnt everywhere and anytime with portable computer to access the sources. This is in compliance with the theories of constructivism and the current learning models. Ubiquitous Learning Environment has the following characteristic [6]: 1) Permanency, 2) Accessibility, 3) Immediacy, 4) Interactivity, 5) Context-awareness, and 6) Adaptability. All of the aforementioned will enable the students to learn anywhere and anytime with an emphasis on context-awareness. Besides, the problem-based learning with mobile device in Ubiquitous Learning Environment will facilitate the learners in solving the real-life problems and communicating with the team members. This will also enhance the efficiency of the learners because there is no limitation of time and place [7],[8].

The instruction that can satisfy the needs of learners and correspond to their contexts is called u-Learning, in which the learners can study anything everywhere and every time through such mobile devices as cellphone or tablet. U-learning also emphasizes on different contexts of learners, such as individuality activity location time and relation [9],[10]. These contexts can be examined and processed to be used as a basis to adjust learning activities, to provide learning contents, and to scaffold the learners during the learning process. Since the learners have to learn by themselves, they need assistance and motivation to learn. The application of u-learning and scaffolding is very suitable to enhance the efficiency of instruction.

Therefore, the researcher had an idea to develop an instructional model that employs the principles of problem-based learning and scaffolding, which correspond to different contexts of learners in ULE. This model will facilitate both instructors and learners. For instance, the learners can study anything everywhere and every time through their mobile devices with certain support systems during the learning process. Thereby, the model will recognize the contexts of learners so that the learners could finally achieve the target of learning. For the instructors, they are able to manage the learning in an efficient manner everywhere and every time. It will be more convenient for them to control the class, check out the learning results, and evaluate the learning.

## 2. PURPOSE OF THE STUDY

**2.1** To develop an Ubiquitous Scaffold Learning Environment Using Problem-based Learning model to enhance problem-solving skills and context awareness.

**2.2** To evaluate the developed Ubiquitous Scaffold Learning Environment Using Problem-based Learning model.

## 3. SCOPE OF STUDY
### 3.1 Population and Sampling Group

**3.1.1** Population of study is experts in the field of instructional design, problem-based learning, scaffolding, ubiquitous learning, information and communication technology and problem solving skills.

**3.1.2** The sample groups are five experts in the field of instructional design, problem-based learning, scaffolding, ubiquitous learning, information and communication technology and problem solving skills, selected by purposive sampling method.

### 3.2 Research Variables

**3.2.1** Independent variable is Ubiquitous Scaffold Learning Environment Using Problem-based Learning model to enhance problem-solving skills and context awareness.
**3.2.2** Dependent variable is evaluation of the model.

## 4. CONCEPTUAL FRAMEWORK

The conceptual framework of this research is to integrate the design of learning of ADDIE Model with the problem-based learning, the scaffolding and Ubiquitous Learning Environment, as shown in Figure 1 [11],[2],[4],[5].

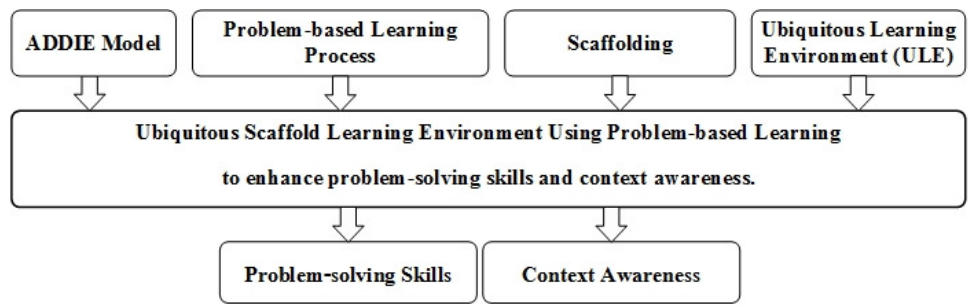

Figure 1. Conceptual Framework

## 5. METHODOLOGY
### 5.1 The First Phase

The development of Ubiquitous Scaffold Learning Environment Using Problem-based Learning model.

**5.1.1** Analyze and synthesize former researches relevant to the elements of problem-based learning, ubiquitous learning environment, scaffolding, problem-solving skills and context awareness.

**5.1.2** Study about learning process by interviewing the instructors in order to synthesize the data of learning activity; and by interviewing the students about their ability to use ICT tools for learning, their learning and cognitive style.

**5.1.3** Design the Ubiquitous Scaffold Learning Environment Using Problem-based Learning model to enhance problem-solving skills and context awareness.

**5.1.4** Present the model to the advisors for consideration and revision.

**5.1.5** Present the model to the experts for consideration by in-depth interview.

**5.1.6** Create the evaluation tools for evaluate the model's suitability.

## 5.2 The Second Phase

This phase was to evaluating the Ubiquitous Scaffold Learning Environment Using Problem-based Learning model.

**5.2.1** Present the designed model to the 5 experts in the field of problem-based learning, ubiquitous learning environment, scaffolding, problem-solving skills and context awareness.

**5.2.2** The model is modified according to the experts' suggestions.

**5.2.3** After modification, presenting the model in the form of diagram with report.

**5.2.4** Analyze the results of evaluation of the model by mean ($\bar{x}$) and standard deviation ($S.D.$) consisting of 5 criteria for evaluation according to the idea of Likert scale.

## 6. RESULTS
### 6.1 Ubiquitous Scaffold Learning Environment Using Problem-based Learning Model

The instructional model is composed of 3 main elements: 1) Elements and principles of the instructional model 2) Objectives of the instructional model 3) Process of the instructional model; the details thereof are shown in Figure 2.

### 6.1.1 Elements and principles of the instructional model include:
### 6.1.1.1 Ubiquitous Learning Environment (ULE)

Ubiquitous Learning Environment in this research is a management of learning environment in order to promote problem-base learning. The said environment consists of 1) Mobile device as a processor: The main devices used in this research are tablet computers or smart phones, 2) Wireless communication technology: It helps learners study and access to the sources with no limit of time and place. The said technology is WiFi that enables mobile devices to connect to the internet, 3) U-LMS: This provides instructional services to store contents of any subjects, learners' data, and learning information, and 4) Context-aware: It is an important part of u-learning and it detects different contexts of learners during the learning process. It also detects learning behaviors of the learners and uses them to produce lesson contents and scaffold the learners.

### 6.1.1.2 Problem-based learning (PBL)

Problem-based learning is a process in u-learning environment, in which the learners will receive an issue from the system to solve. The learners can solve the said problem anywhere any time. They will also be scaffolded according to their learning contexts; this will assist them to succeed in solving that problem. Learners will receive and solve a problem situation as assigned by the system. They could work on it anywhere and anytime. The problem-based learning comprises seven steps as follows [4][7]: 1) present the problem; 2) clarify concept and definition of the problem; 3) development of hypotheses and their sequence; 4) formulate learning objectives; 5) collect additional information; 6) synthesize and test newly acquired information; and 7) reflect and evaluate. Through all these steps, the scaffolding is automatically adjusted and provided in accordance with learners' context by referring to the context awareness attribute of Ubiquitous Learning Environment so that learners could solve the problem.

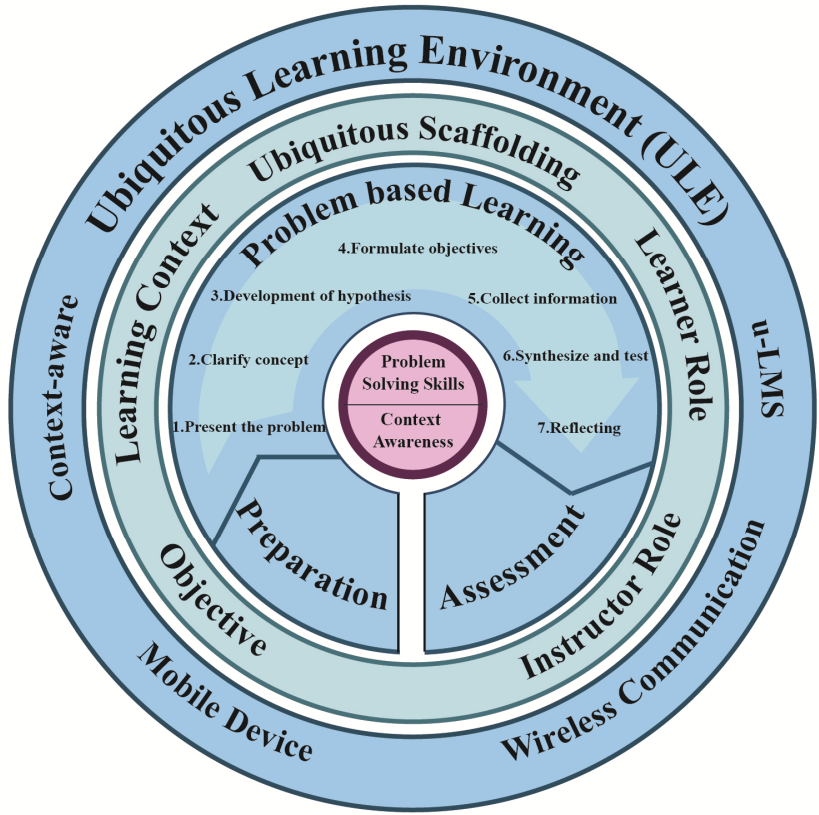

Figure 2. Ubiquitous Scaffold Learning Environment Using Problem-based Learning Model

### 6.1.1.3 u-Scaffolding

This scaffolding part can help the learners solve some problems that they cannot do by themselves in the process. This scaffolding part relies on the advantage of Ubiquitous learning; i.e. Context-awareness. Thereby, the assistance is based on learning contexts of learners including: 1)individuality: contexts about personal information of learners 2) activity: instructional activities such as learning results, test results, activity clues 3) location: the locations of learners 4) time: learning duration of students, appointments 5) relation: contexts about the relation between learners and friends, e.g. interaction between learners' group and instructors. These contexts are detected by Learner' Context-aware Module so as to provide different kinds of assistance such as 1) Reminder: to notify about the activity to do 2) Coaching: to provide the learners with further information so that they can learn more and understand more, and 3) Mentor arrangement: The system will match the learner with a partner for conversation, who can offer helpful advice to find out the solutions.

### 6.1.1.4 Roles of instructors

The instructors are required to manage and divide the learners into groups, hold an orientation to introduce u-learning, show how to use mobile devices, manage learning and media contents corresponding to the subjects, set up instructional activity including both learning and activity schedules, evaluate learning results and send feedback to the learners so that the latter could see their progress, follow up and facilitate the learners in any activity, and provide some advice when requested by the learners.

### 6.1.1.5 Roles of learners

The learners do their learning activity as provided by the instructors. Thereby, the learners can learn anything anywhere and anytime through mobile devices. At the time of learning, the system will notify the learners via their mobile device about the upcoming activity. During this process, the learners will receive assistance automatically from Learner' Context-aware Module. For instance, they will be provided with additional learning contents, consults from friends and instructors. Meanwhile, the learners are able to communicate with their friends and instructors all the time.

### 6.1.1.6 Problem Solving Skill.

Problem Solving Skill. Means the ability to reflect, collect, analyzes and assesses information for making a decision to suggest a method by which obstacles or undesired situation are eliminated or diminished. According to Weir [12], it comprises four sub skills: 1) problem identification; 2) cause-problem analysis; 3) problem solution; and 4) result evaluation.

### 6.1.1.7 Context awareness

Context awareness means the ability of instructional system to detect different contexts of learners during the process. This is to encourage the learners to achieve their learning. The context awareness has the following characteristics [9][10].

1) Individuality: aware of learners' characteristics such as personal information, interest, and ability of learners.
2) Activity: aware of instructional activities of learners and management of their learning trends.
3) Location: aware of locations of learners who are in the process of learning so that their location information would be used to analyze the behaviors of learners.
4) Time: aware of instructional schedules in order to notify the learners at the time of activity
5) Relation: aware of the social relation between learners, e.g. social activity of learners, assistance from classmates or from instructors.

### 6.1.2 The objective of instructional model.

The objective of this instructional model is to enhance problem-solving skills and context awareness.

### 6.1.3 Process of the instructional model.

Process of the instructional model can be divided in three stages:

### 6.1.3.1 Preparation stage

1) Orientation. Instructors explain details and provide guidance about the ubiquitous learning.
2) Registration and workshop. Instructors distribute tablet computers to all learners for basic use workshop. Learners register, test the system access, and practice using the learning management system for each subjects, discussion board, and assignment submission.

3) Learner grouping. Groups of five members are formed on voluntary basis and each member's role is identified. The information about each group is then posted via u-LMS interface.

4) Evaluation of Problem-solving skill. It is executed before learners commence the learning.

### 6.1.3.2 Learning stage

The instructional process is in the format of Ubiquitous learning environment, in which the learners will study online wherever there is WiFi, e.g. home, university, or public areas with WiFi service, with no need to come to class. The learning is based on problem-based learning together with scaffolding system in each step of problem solving. This can scaffold the learners and detect different contexts of learners. Once the system is aware of learners' contexts, it will respond in the form of scaffolding. This includes 1) Reminder: to automatically notify the learners about the activity to do and lesson contents, 2) Coaching: to teach something in addition to those in the system depending on weekly test results and duration spent on each learning step of problem solving (In the first week, all learners receive scaffolding so that they could achieve the goals of each learning step. However, for the learning of the following week, the assistance is based on the scores derived from each step of the previous week. That is if the scores do not meet the criteria, there is still assistance from the system. In contrast, if the scores meet the criteria, the system will stop offering assistance, and 3) Mentor arrangement: the system will introduce the learners to talk to the classmates who have high scores so that the latter can offer helpful advice to find out the solutions. Details of each learning steps are described as follows:

1) Presentation of problems. When a study time comes, learners receive a reminder message from the system sent to their tablet computer so they get ready for the forthcoming learning. The system then sends a problem situation to learners automatically.

2) Problem understanding and identification. Learners study and understand the problem situation received from the system through discussion and brainstorming among their group members via u-LMS interface on the tablet computer. Then learners submit the result of problem identification to the system.

3) Development of hypotheses and their sequence. Group members undertake discussion and brainstorming to analyze causes of the problem and develop hypotheses. Members can express their views freely. Learners sequence the hypotheses, write them down and submit to the system.

4) Formulation of learning objectives. Group discussion and brainstorming are undertaken to formulate learning objectives that give direction for the search of additional information to prove the selected hypotheses. Learners summarize the result of learning objective formulation and submit it to the system.

5) Collection of additional information. Members divide tasks of additional data research from knowledge sources. The sought data shall correspond with the formulated learning objectives and be looked up on the tablet computer.

6) Synthesis and test of the newly acquired information. Members jointly produce a summary of the research and separate the hypothesis-supporting information from that non-hypothesis-supporting. Together they make a conclusion of the supporting information and test if it is sufficient to prove the hypothesis. Learners submit the result of the information synthesis and test.

7) Learning conclusion. Group members discuss and brainstorm via chat room and produce a conclusion on the learning, principles, and ideas obtained through the problem-based study. Learners input the result reflecting their own view in the system.

### 6.1.3.3 Evaluation stage.

1) Evaluation of problem-solving skill. After finishing the study, learners' problem-solving skill is evaluated by using closed form questionnaire about the problem-solving skill. Four aspects of the skill are evaluated: 1) problem identification; 2) cause-problem analysis; 3) problem solution; and 4) result evaluation.

2) Evaluation of context awareness. Evaluation of context awareness is an assessment of the model's characteristics that are aware of learners's contexts. The assessment is done by experts and learners; whereby the characteristics to be appraised are individuality activity location time and relation.

**6.2 Results of evaluation of model suitability**

Evaluation of model suitability was conducted by five experts and the results thereof were presented in Table 1-4.

Table 1. Evaluation of the overall elements of this instructional model.

| Evaluation Lists | Results | | Appropriateness |
|---|---|---|---|
| | $\bar{x}$ | S.D. | |
| 1. Components and principles of the model. | 4.80 | 0.45 | Highest |
| 2. Objectives of the learning model. | 4.40 | 0.89 | High |
| 3. Learning process. | 3.60 | 1.14 | High |
| 4. Evaluation | 4.50 | 0.58 | Highest |
| **Summary** | **4.33** | **0.76** | **High** |

Table 1: The experts found that the overall elements of this instructional model were suitable at high level ($\bar{x}$ = 4.33, S.D. = 0.76). This means the elements and the process of this instructional model were suitable to be learning environment.

Table 2. Evaluation of preparation stage before learning.

| Evaluation Lists | Results | | Appropriateness |
|---|---|---|---|
| | $\bar{x}$ | S.D. | |
| 1. Orientation | 4.80 | 0.45 | Highest |
| 2. Registration and workshop | 4.60 | 0.55 | Highest |
| 3. Learner grouping | 4.60 | 0.55 | Highest |
| 4. Evaluation of Problem-solving skill before learning | 5.00 | 0.00 | Highest |
| **Summary** | **4.75** | **0.39** | **Highest** |

Table 2: The experts found that the preparation step before learning was suitable at highest level ($\bar{x}$ = 4.75, S.D. = 0.39). It means this step could equip the learners with different preparedness so that they could do well in the instructional management step.

Table 3. Evaluation of learning stage.

| Evaluation Lists | Results | | Appropriateness |
|---|---|---|---|
| | $\bar{x}$ | S.D. | |
| 1. Present the problem. | 5.00 | 0.00 | Highest |
| 2. Clarify concept and definition of the problem. | 4.80 | 0.45 | Highest |
| 3. Development of hypotheses and their sequence. | 4.60 | 0.89 | Highest |
| 4. Formulate learning objectives. | 4.80 | 0.45 | Highest |

| Evaluation Lists | Results | | Appropriateness |
|---|---|---|---|
| | $\bar{x}$ | S.D. | |
| 5. Collect additional information. | 4.80 | 0.45 | Highest |
| 6. Synthesize and test newly acquired information. | 5.00 | 0.00 | Highest |
| 7. Reflect and evaluate | 4.80 | 0.45 | Highest |
| **Summary** | **4.83** | **0.38** | **Highest** |

Table 3: The experts found that the learning stage was suitable at highest level ($\bar{x}$ = 4.83, S.D. = 0.38). This means the instructional management herein was suitable for instructional activity.

Table 4. The evaluation results of application of this instructional model.

| Evaluation Lists | Results | | Appropriateness |
|---|---|---|---|
| | $\bar{x}$ | S.D. | |
| 1. The model is appropriate to enhance problem-solving skill. | 4.80 | 0.45 | Highest |
| 2. The model is appropriate to enhance context awareness. | 4.60 | 0.55 | High |
| 3. The model is possible for using. | 4.60 | 0.55 | High |
| **Summary** | **4.67** | **0.52** | **High** |

Table 4: The experts found that the application of this instructional model was suitable at high level ($\bar{x}$ = 4.67, S.D. = 0.52). This means the model is suitable for the development of problem-solving skills and it is really practical.

## 7. DISCUSSION

According to the results of this research, the discussion is as below:

**7.1** According to the evaluation, the overall elements of this instructional model is suitable at high level. This is because the development of this model applied the concepts of Ubiquitous learning environment with the problem-based learning. This corresponds to the research of Jones and Jo [5] and Ogata [8], who found that this kind of learning environment would enable the learners to study everywhere at any time, complying with the theories of Constructivism.

**7.2** According to the evaluation, the preparation stage before learning was suitable at highest level. It is important to make the learners well prepared in every aspect so that the learners could learn anything based on the model as efficiently as possible. This is in accordance to Oldham et al [13], who suggested that the learners should practice using different technology in order to have experience and preparedness for learning.

**7.3** According to the evaluation, the learning stage was suitable at very high level. This is because the application of problem-based learning process can enhance the problem-solving skills of learners. Besides, since the Learners' Context-aware Module can detect the learners' contexts, the instructional activity can be adapted to increase the learners' success in learning [10].

**7.4** According to the evaluation, application of this instructional model was suitable at high level. This model can develop the problem-solving skills and use u-learning model. Thanks to the present technology of such equipment as tablet computer and various mobile devices, it is possible to create learning environment based on the said model [14].

## 8. SUGGESTIONS

Suggestions on application of this research's results

**8.1** Any education institutes that apply this instructional model should prepare their infrastructure, and prepare both learners and instructors so that they could create Ubiquitous learning environment.

**8.2** The developed model can be applied with undergraduates of all field and all levels. Also, this model can be used in all theoretical sections of all subjects.

**8.3** There should be motivations for the learners during learning process in order to stimulate their enthusiasm to learn.

**8.4** The limitation of using this model is that the learners must have mobile devices with internet connection such as tablet or cell phone.

## 9. FURTHER RESEARCH

The results of this research should be used and tested to see the results of learning based on this model, e.g. learning achievement, problem-solving skills, or attitudes of learners towards the instructional model.

**Authors**

**Noppadon Phumeechanya** is a Ph.D. candidate, Department of Information and Communication Technology for education, Faculty of Technical Education, King Mongkut's University of Technology North Bangkok (KMUTNB) and He is lecturer at Department of Computer and Information Technology, Faculty of Science and Technology, Nakhon Pathom Rajabhat University, Thailand. He has experience in Information and Communication Technology for Education research.

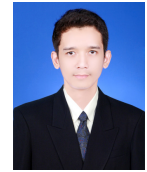

**Dr.Panita Wannapiroon** is an Assistant Professor at Division of Information and Communication Technology for Education, Faculty of Technical Education, King Mongkut's University of Technology North Bangkok (KMUTNB),Thailand.
She has experience in many positions such as the Director at Innovation and Technology Management Research Center, Assistant Director of Online Learning Research Center, Assistant Director of Vocational Education Technology Research Center, and Assistant Director of Information and Communication Technology in Education Research Center. She received Burapha University Thesis Award 2002. She is a Membership of Professional Societies in ALCoB (APEC LEARNING COMMUNITY BUILDERS) THAILAND, and Association for Educational Technology of Thailand (AETT)

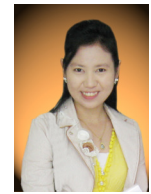